\documentclass[twocolumn,showpacs,aps,prb,letterpaper,nofootinbib,raggedbottom,nobalancelastpage]{revtex4}
\usepackage{graphicx,bm}
\usepackage{amsmath}
\usepackage{color}

\begin{document}
\title{Generation of time-bin entangled photon pairs using a quantum-dot cavity system}
\date{\today}
\author{P. K. Pathak and S. Hughes}
\address{Department of Physics,
Queen's University, Kingston, ON K7L 3N6, Canada}
\begin{abstract}
We present a scheme to realize a deterministic solid state source of
time-bin entangled photon pairs using cavity-assisted  stimulated
Raman adiabatic passage (STIRAP) in a single quantum dot. The
quantum dot is embedded inside a semiconductor cavity, and the interaction of a coherent
superposition of two temporally separated input pulses and the cavity
mode leads to a two-photon Raman transition, which produces a time-bin
entangled photon pair  through the  biexciton-exciton cascade. We show that the entanglement of
the generated state can be measured using triple coincidence
detection, and the degree of entanglement is quantified as the visibility of the interference.
We also discuss the effect of pure dephasing on
entanglement of the generated photon pair. Pronounced interference
visibility values of greater than $1/\sqrt{2}$ are demonstrated in
triple coincidence measurement using experimentally achievable
parameters, thus demonstrating that the generated photons are suitable for applications
with Bell's inequality violation and quantum cryptography.
\end{abstract}
 \maketitle

\section{Introduction}
A source of entangled photon pairs is an essential building block
for various quantum information processing protocols\cite{qinfo}, such as
quantum cryptography\cite{crypto} and quantum teleportation\cite{tele}. Generally, the
employed entangled state of photons in these experiments are
entangled in both the energy and the polarization degrees of freedom\cite{kwait}. However,
because of unavoidable polarization dispersion in optical fibers, the
polarization entangled photons are not suitable for distribution
over  large distances. In related experiments\cite{tbin,tbin2,tbin3,tbin4},
entangled states of photons in energy and time degrees of freedom,
using discrete time interval (time-bin) for photon emission, have
been demonstrated and the entanglement between these photons has
been successfully distributed over distance of 50\,km \cite{tbin4}. In these
experiments, the time-bin entangled photons were  generated through
the parametric down convertor (PDC) using the pump as a
superposition of two time separated pulses. A PDC is a heralded
source of entangled photons where the number of generated photon
pairs are probabilistic \cite{mandelwolf}. At low pump intensity, when the probability
of generating more than one photon pair remains small, the
efficiency of the source remains very low (less than 20\%
\cite{tbin3}). For quantum information processing applications, one
requires a scalable source which generates precisely a single photon
pair on demand\cite{scalable}. In the last few years, there has been considerable
progress for developing {\em on demand} single photon and entangled photon
sources using single quantum dots (QDs)\cite{yamamoto,akopian,mfield,efield}, where the QDs provide the
potential advantages of integrability and scalability in such
experiments. In semiconductor QDs, polarization entangled photons have been
successfully generated in the biexciton-exciton cascade decay \cite{akopian,mfield,efield}.

In 2005, Simon and Poizat \cite{simon} proposed an on-demand
generation of time-bin entangled photons through the
biexciton-exciton cascade in idealized QDs, where the bi-exciton state is created
by pumping through two pulses interacting at two distinct times. The
state of the time-bin entangled photon pair is given by $
|\psi\rangle=\sqrt{p_1}|early\rangle_1|early\rangle_2+e^{i\theta}\sqrt{p_2(1-p_1)}|late\rangle_1|late\rangle_2,
$ where {\em early} and {\em late} are two time bins and $p_1$ is the
probability of generating a photon pair in the early time bin (from the first
pulse) and $p_2$ is probability of generating a photon pair in the late
time bin (from the second pulse); the total probability is then $p_1+p_2=1$. For generating maximally entangled state $|\psi\rangle$, one requires $p_1=p_2=1/2$. Therefore, a precisely regulated  population transfer between the QD energy levels is essential. Moreover, pure dephasing processes present in semiconductor produce detrimental effects on the entanglement of the generated state.
In quantum information protocols,
such as entanglement swapping, it is essential that the photons should not have any other correlation
except the time-bin entanglement. However, in the biexciton-exciton cascade, emitted photons also have time correlations.
These undesirable temporal correlations can be minimized by manipulating emission rates of photons using resonant cavities\cite{simon}.

In this work, we propose to generate an efficient time-bin entangled photon pair using
stimulated Raman adiabatic passage (STIRAP). The coherent excitation
in the system of QDs embedded in a semiconductor cavity have been
an active area of research \cite{coherent,ates}.
We consider the intitial QD state is in a metastable state,
and there have been several methods for achieving this using
electrical control of QD-cavity mode resonce\cite{jelena,electric}.
 We demonstrate that the STIRAP process then provides an efficient regulated way for population transfer.
 We also investigate how the cavity enhanced decay rates suppress
 the detrimental effects of pure dephasing.

Our paper is organized as follows. In Sec.~II, we present a formal
theory of generation of time-bin entangled photon pair from a single
QD coupled to a semiconductor cavity. In Se.~III, we investigate
the measure of photon entanglement by a triple coincidence detection and also study the effects
of dephasing. In section~IV, we  present our conclusions.

\section{Generating time-bin entangled photon pairs using STIRAP}
\label{Sec:Theory} 
\begin{figure}[t!]
\centering
\includegraphics[width=3in,height=2.5in]{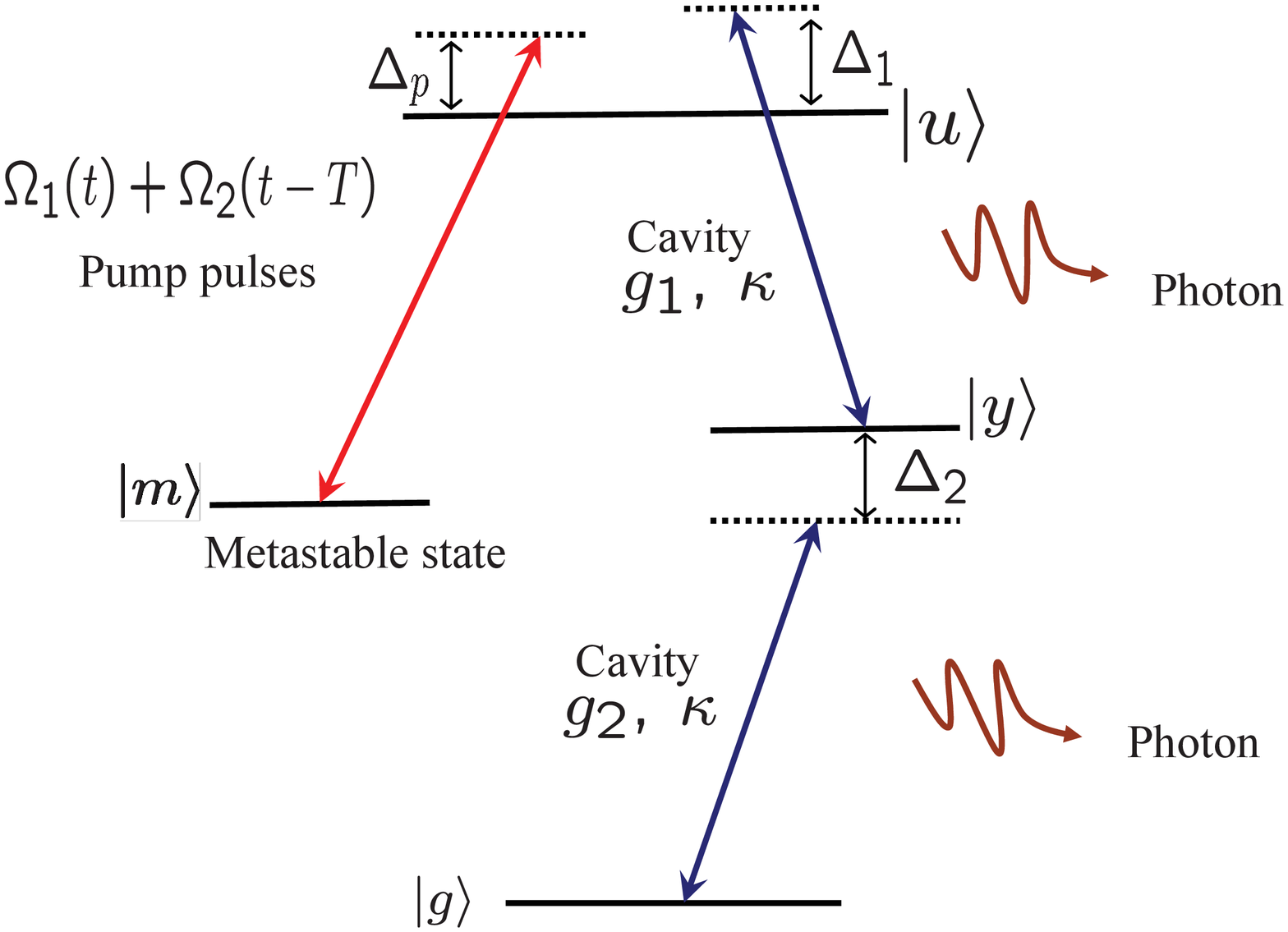}
\vspace{-0.5cm}
\caption{(Color online) Schematic diagram for time-bin entangled photon generation. The
transitions from the biexciton state $|u\rangle$ to the exciton
states $|y\rangle$  and $|y\rangle$ to ground state $|g\rangle$ are coupled by a $y$-polarized cavity mode.
The QD is pumped from the metastable state $|m\rangle$ to biexciton state through the superposition of two input pulses.} \label{fig1}
\end{figure}

We consider a QD embedded in a semiconductor
microcavity, where the energy level diagram of the system is shown in Fig.~\ref{fig1}. The dipole transitions
from the biexciton state $|u\rangle$ to the exciton state $|y\rangle$,
and from the exciton state $|y\rangle$ to the ground state $|g\rangle$, are
coupled through a $y$-polarized single mode of the semiconductor
cavity, with coupling constants $g_1$ and $g_2$, respectively.
Because of the large biexciton binding energy of semiconductor
QDs, it is not
possible to couple the biexciton and exciton transitions from the
same cavity mode, and thus manipulation of the biexciton binding energy
becomes essential in these systems\cite{ourprb,reimer}. Usually the
binding energy of the charge-neutral biexciton has a negative value, however by changing the
confinement size\cite{size} or by changing the
strain\cite{strain}, it has been found that the biexciton
binding energy can be tuned to zero or a positive value.
Very recently, manipulation of the binding energy of the biexciton
has also been reported by applying lateral electric
fields\cite{reimer}. Moreover, construction of an
electrode for applying a lateral electric field in the vicinity of a
QD within a photonic crystal cavity has also been
reported\cite{jelena}. Therefore, it is now possible to manipulate
the binding energy of biexcitons inside semiconductor cavities.
The another advantage of lateral electric fields is that they can
be used to create a voltage-tunable metastable exciton state
$|m\rangle$.

In III-V QDs,  bright netral excitons
are formed when an electron from the $s-$shell or the $p-$shell of the valance
band is excited to the $s-$shell or $p-$shell of the conduction band,
respectively; the transitions from $s-$shell to $p-$shell and vice versa
are essentially {\em symmetrically forbidden}. However, in the presence of an applied lateral
voltage,
the charge carrier symmetry can be suitably broken broken so that
a bright exciton is formed, e.g.,  from the  $s-$shell conduction band to
$p-$shell conduction band\cite{reimer}. Consequently, by applying a lateral voltage larger
than the values required to break the symmetry, and then using a
$\pi$ pulse excitation to this state, a symmetrically forbidden exciton can be
created which behaves as the metastable state $|m\rangle$, after lowering the applied voltage again.
In what follows below, we will assume that this initial state can be created
and focus on its evolution after applying STIRAP pulses.

Initially, the QD is prepared in the metastable state $|m\rangle$
and the cavity mode in the vacuum state. An $x$-polarized pump field
with a Rabi frequency $\Omega_p(t)$ is applied between the metastable
state $|m\rangle$ and the biexciton state $|u\rangle$. The
Hamiltonian of the system, in the rotating frame at the pump
frequency (interaction picture),  can be written as
\begin{eqnarray}
&&H  = \hbar\Delta_p|m\rangle\langle m|+\hbar\Delta_1|y\rangle\langle y|+\hbar(\Delta_1+\Delta_2)|g\rangle\langle g|\nonumber\\
&&+   \hbar\left[\Omega_p(t)|u\rangle\langle m|
+g_1|u\rangle\langle y|a+g_2|y\rangle\langle g|a+H.c.\right],
\label{ham}
\end{eqnarray}
where $\Delta_p$, and $\Delta_1$ ($\Delta_2$) are the detunings of
the pump field, and the cavity mode with the biexciton (exciton)
transition frequencies, respectively;
 $H.c$ refers to Hermitian conjugate. For simulating the dynamics of the system,
we perform quantum master equation calculations in the density matrix
representation.
 The evolution of the QD-cavity system is given by
\begin{eqnarray}
{\dot\rho}=-\frac{i}{\hbar}[H,\rho]-\frac{1}{2}  \sum_{\mu} [L_{\mu}^{\dag}L_{\mu}\rho
-2L_{\mu}\rho L_{\mu}^{\dag}+\rho L_{\mu}^{\dag}L_{\mu}] ,
\label{master}
\end{eqnarray}
where $L_{\mu}$ are the Lindblad operators, with terms
$\sqrt{\gamma_1}|m\rangle\langle u|$,
$\sqrt{\gamma_1}|y\rangle\langle u|$, and
$\sqrt{\gamma_2}|g\rangle\langle y|$ corresponding to the
spontaneous decays, and $\sqrt{2\gamma_d}|u\rangle\langle u|$,
$\sqrt{\gamma_d}|m\rangle\langle m|$, and
$\sqrt{\gamma_d}|y\rangle\langle y|$, corresponding to pure dephasing
of the biexciton and exciton states. The emission of the photons from
the cavity mode is given by the Lindblad operator $\sqrt{\kappa}a$,
where $2\kappa$ is the decay rate of the leaky cavity.
We  safely neglect the spontaneous decay of the metastable state $|m\rangle$
during the evolution, as the lifetime of the metastable state is, by definition, very
large.

We numerically solve the optical Bloch equations using Eq.~(\ref{master}), for
density matrix elements $\rho_{ij}=\langle i|\rho|j\rangle$.
To simplify the nottaion, we use the
definitions $|Y\rangle=|y,1\rangle$, $|G\rangle=|g,1\rangle$, and
$|G'\rangle=|g,2\rangle$, where the alpha-numeric notation corresponds to the energy state of QD
 and the number corresponds to the cavity photon. The complete dynamics of the
 system is expressed by the following
equations of motion:
\begin{eqnarray}
\label{meq1}
\dot\rho_{mm}&=&-i\Omega^*_p(t)\rho_{um}+i\Omega_p(t)\rho_{mu}+\gamma_1\rho_{uu},\\
\dot\rho_{uu}&=&i\Omega^*_p(t)\rho_{um}-i\Omega_p(t)\rho_{mu}+ig_1\rho_{uY}-ig_1\rho_{Yu} \\ \nonumber
&& -2\gamma_1\rho_{uu},\\
\dot\rho_{YY}&=&-ig_1\rho_{uY}+ig_1\rho_{Yu}-ig_2\sqrt{2}\rho_{G'Y}+ig_2\sqrt{2}\rho_{YG'}\nonumber\\
&&-(\kappa+\gamma_2)\rho_{YY},\\
\dot\rho_{G'G'}&=&ig_2\sqrt{2}\rho_{G'Y}-ig_2\sqrt{2}\rho_{YG'}-2\kappa\rho_{G'G'},\\
\dot\rho_{GG}&=&\!ig_2(\rho_{Gy}\!+\!\rho_{yG})+2\kappa\rho_{G'G'}+\gamma_2\rho_{YY}-\kappa\rho_{GG}, \ \\
\dot\rho_{yy}&=&-ig_2(\rho_{Gy}\!-\!\rho_{yG})+\kappa\rho_{YY}+\gamma_1\rho_{uu}-\gamma_2\rho_{yy}, \ \\
\dot\rho_{gg}&=&\kappa\rho_{GG}+\gamma_2\rho_{yy},\\
\dot\rho_{um}&=&-i\Delta_p\rho_{um}-i\Omega_p\rho_{mm}-ig_1\rho_{Ym}+i\Omega_p\rho_{uu}\\
&&-(\gamma_1+\frac{3\gamma_d}{2})\rho_{um},\\
\dot\rho_{uY}&=&-i\Delta_1\rho_{uY}-i\Omega_p\rho_{mY}-ig_1\rho_{YY}+ig_1\rho_{uu}\nonumber\\
&&+ig_2\sqrt{2}\rho_{uG'}-(\gamma_1+\frac{\kappa+\gamma_2+3\gamma_d}{2})\rho_{um},\\
\dot\rho_{YG'}&=&-i\Delta_2\rho_{YG'}-ig_1\rho_{uG'}-ig_2\sqrt{2}\rho_{G'G'}\nonumber\\
&&+ig_2\sqrt{2}\rho_{YY}-(\frac{3\kappa+\gamma_2+\gamma_d}{2})\rho_{YG'},\\
\dot\rho_{mY}&=&-i(\Delta_1-\Delta_p)\rho_{mY}-i\Omega^*_p(t)\rho_{uY}+ig_1\rho_{mu}\nonumber\\
&&+ig_2\sqrt{2}\rho_{mG'}-(\gamma_d+\frac{\kappa+\gamma_2}{2})\rho_{mY},\\
\dot\rho_{uG'}&=&-i(\Delta_1+\Delta_2)\rho_{uG'}-i\Omega_p(t)\rho_{mG'}-ig_1\rho_{YG'}\nonumber\\
&&+ig_2\sqrt{2}\rho_{uY}-(\gamma_d+\kappa+\gamma_{2})\rho_{uG'},\\
\dot\rho_{mG'}&=&-i(\Delta_1+\Delta_2-\Delta_p)\rho_{mG'}-i\Omega^*_p(t)\rho_{uG'}\nonumber\\
&&+ig_2\sqrt{2}\rho_{mY}-(\kappa+\frac{\gamma_d}{2})\rho_{mG'},\\
\dot\rho_{yG}&=&-i\Delta_2\rho_{yG}-ig_2\rho_{GG}+ig_2\rho_{yy}\nonumber\\
&& -\frac{(\kappa+\gamma_2+\gamma_d)}{2}\rho_{yG},
\label{meq2}
\end{eqnarray}

\begin{figure}
\vspace{-0.4cm} \centering
\includegraphics[width=3in,height=2.5in]{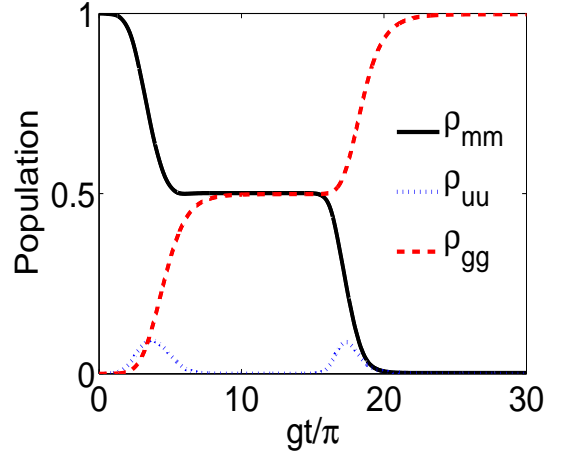}
\caption{(Color online)Populations in state $|m\rangle$ (black solid line), in state
$|g\rangle$ (red dashed line) and in state $|u\rangle$ (blue dotted line) during the evolution.
Two Gaussian pump pulses of different amplitude are employed such
that $p_1=p_2=1/2$. Using the parameters $g_1=g_2=g$, $\kappa=2.5g$, $\gamma_2=\gamma_1=10^{-3}g$, $\gamma_d=10^{-2}g$, $\Delta_1=-3g$, $\Delta_2=2g$, $\Delta_p=-1.5g$, $\Omega_1(t)/g=0.74\exp[-(t-2\tau_p)^2/\tau_p^2]$, $\Omega_2(t)/g=3\exp[-(t-9.5\tau_p)^2/\tau_p^2]$, and $g\tau_p=2\pi$.} \label{fig2}
\end{figure}

When the pump field and the cavity mode satisfy the two-photon Raman
resonance condition $\Delta_p\approx\Delta_1+\Delta_2$, the evolution of the population in the QD
energy levels follows the cavity-assisted STIRAP.
Here, one must also remember
that the Stark shifts of
energy levels also play an important role in two-photon Raman
resonance condition\cite{stirap}. However, for constant cavity
couplings, the Stark shifts in energy states remains constant and the
two-photon Raman resonance condition can be satisfied easily by changing
the detuning of the pump field ($\Delta_p$) only.

In Fig.~\ref{fig2},
we show the numerical simulations after solving Eqns.~(\ref{meq1})-(\ref{meq2}).
The pump field is chosen to be a coherent
superposition of two time-separated Gaussian pulses of the same width,
but different amplitudes, which can be generated by passing a
Gaussian pulse through an unbalanced two arm interferometer. The
Rabi frequency of the pump  field is given by
$\Omega_p(t)=\Omega_1(t)+\Omega_2(t-T)$, where $\Omega_{1,2}(t)$ is
the Rabi frequency of each pulse and $T$ is the time gap between the
pulses.  We select a typical value of $\Omega_1(t)$ such that the
population of the state $|m\rangle$ is pumped to the state $|y,1\rangle$ in
STIRAP with probability $p_1=1/2$. Due to the nature of the leaky cavity mode, the
photon is emitted from the final state $|y,1\rangle$ and the system
is evolved into the state $|y,0\rangle$ state. The population in
state $|y,0\rangle$ is transferred to the state $|g,1\rangle$
through the cavity mode. After emitting another photon from the
state $|g,1\rangle$, the system finally reaches the ground state
$|g,0\rangle$. Thus a photon pair in the early time bin is emitted
during the interaction of the first pulse $\Omega_1(t)$. The
remaining population in state $|m\rangle$ is similarly pumped by
another pulse $\Omega_2(t-T)$ and a photon pair is generated with
probability $p_2=1-p_1$ in the late time bin. The state of the
generated photon pair is thus maximally time-bin entangled state. For
a QD embedded in a microcavity, the off-resonant exciton has
a spontaneous decay rate of the order of $0.1-1\,\mu$eV ($10^{-3} g$ for
 $g =0.1\,$meV), and  the cavity decay condition $\kappa>>g$ can be achieved easily.
 We stress that all of these parameters correspond
 closely to those in present day experiments~\cite{coherent,ates}.

During the time interaction with the pump pulses, the population in the upper state $|u\rangle$ remains
always less that $0.1$ and the population in $|g,2\rangle$ remains
negligible. The state of emitted photon pair from the
cavity mode can therefore be written as
\begin{equation}
|\psi(t)\rangle=\left[a^{\dag}_1(t)a^{\dag}_2(t)+a^{\dag}_1(t-T)a^{\dag}_2(t-T)\right]|\{0\}\rangle,
\label{tstate}
\end{equation}
where $|\{0\}\rangle$ is vacuum field, $\langle
a_1(t)\rangle=\langle\sigma_{yY}(t)\rangle$,
and $\langle a_2(t)\rangle=\langle\sigma_{gG}(t)\rangle$, for
$\sigma_{ij}=|i\rangle\langle j|$. In quantum information protocols,
such as entanglement swapping, it is essential that the photons in
the mode $a_1$ and $a_2$ should not have any other correlation
except the time-bin entanglement. However, in the biexciton-exciton cascade,
the $a_1$ mode photon is always generated after the emission of
the $a_2$ mode photon. Thus the $a_1$ and $a_2$ modes remain time correlated.
This undesirable temporal correlation becomes negligible for
$\Gamma_1/\Gamma_2>>1$ \cite{simon}, where $\Gamma_i$ is the emission
rate of the photon in $a_i$ mode. In our scheme above, the first photon in mode $a_1$ is generated in resonant Raman process, which is emitted with the cavity mode decay rate $\kappa$, and the second photon is generated through cavity enhanced spontaneous emission. The condition $\Gamma_1/\Gamma_2>>1$ can
therefore be easily satisfied by choosing a suitably  large value of the detuning
$\Delta_2$ such that $g_2^2\kappa/(\kappa^2+\Delta_2^2)<<\kappa$.

\section{Triple coincident detection of the photon entanglement, and the influence of pure  dephasing}
Next, we discuss how to measure the entanglement of the generated
state of the photons. The Concurrence of the state (\ref{tstate}) is
directly related to the coherence of the state \cite{akopian}. For
measuring the degree of entanglement, photons from each mode are passed through
an unbalanced two path interferometer\cite{tbin}; the time difference between
two arms is $T$, with phase difference $\phi$,  and $T$ is similar to the time difference between the two pulses in the
pump fields.
After passing through the
interferometers, the field operators at the output of the interferometers can be expressed as
\begin{eqnarray}
\label{a3} a_3(t) =a_1(t)+e^{i\phi}a_1(t-T), \ \ \\
a_4(t)=a_2(t)+e^{i\phi}a_2(t-T). \ \ \label{a4}
\end{eqnarray}
The post-selection, for detecting both photon simultaneously after passing through the interferometers, projects the state (\ref{tstate}) into the state
\begin{eqnarray}
|\psi_c(t)\rangle=\left[a^{\dag}_1(t)a^{\dag}_2(t)+(1+e^{2i\phi})a^{\dag}_1(t-T)a^{\dag}_2(t-T)\right.\nonumber\\
\left.+a^{\dag}_1(t-2T)a^{\dag}_2(t-2T)\right]|\{0\}\rangle. \
\label{post}
\end{eqnarray}
Clearly, the state (\ref{post}) has three terms which are distinguishable in time. The middle term, appearing at $t=T$, provides the information about the entanglement of state (\ref{tstate}). For separating different terms in state (\ref{post}), the time of detection of photons is measured with reference to the pump photons using a triple coincidence detection.
The probability of triple coincidence detection of one
photon at the output of each interferometer, and one from the input pulse $\Omega_1$,
is given by
\begin{eqnarray}
&&\!\!G^{(3)}(\tau)=\int_0^{\infty} dt'\int_{-T_{bin}}^{T_{bin}}d\tau'
|\Omega_1(t')|^2 \times \nonumber\\
&&\!\!\langle
a^{\dag}_3(t'+\tau)a^{\dag}_4(t'+\tau+\tau')a_4(t'+\tau+\tau')a_3(t'+\tau)\rangle,
\label{corr} \ \ \ \ \
\end{eqnarray}
where $T_{bin}$ is the width of the time-bins; which is chosen larger than the biexciton-exciton cascade decay and smaller than $T$.
We can simplify the above expression for $G^{(3)}(\tau)$ using the property of field operators, $a_1(t)a_2(t-T)|\{0\}\rangle=0$, as both photons are generated almost together in the cascade decay.
Subsequently, we can simplify the correlation function in (\ref{corr}) as
%
\begin{widetext}
\begin{eqnarray}
\langle a^{\dag}_3(t'+\tau)a^{\dag}_4(t'+\tau+\tau')a_4(t'+\tau+\tau')a_3(t'+\tau)\rangle&&
 \!\!\!\!=  \langle a^{\dag}_1(t'+\tau)a^{\dag}_2(t'+\tau+\tau')a_2(t'+\tau+\tau')a_1(t'+\tau)\rangle  \nonumber \\
&\hbox{} &
\!\!\!\! + \ \langle a^{\dag}_1(t'-T+\tau)a^{\dag}_2(t'-T+\tau+\tau')a_2(t'-T+\tau+\tau')a_1(t'-T+\tau)\rangle \nonumber \\
&\hbox{} &
\!\!\!\! +  \ 2\cos2\phi\langle a^{\dag}_1(t'+\tau)a^{\dag}_2(t'+\tau+\tau')a_2(t'-T+\tau+\tau')a_1(t'-T+\tau)\rangle, \ \ \ \ \ \
\end{eqnarray}
\end{widetext}
which is evaluated for state (\ref{tstate}) by applying
quantum regression formula\cite{book}. We relegate the details
of the $G^{(3)}$ calculation to the Appendix.

\begin{figure}
\vspace{-0.4cm}\centering
\includegraphics[width=3in,height=2.5in]{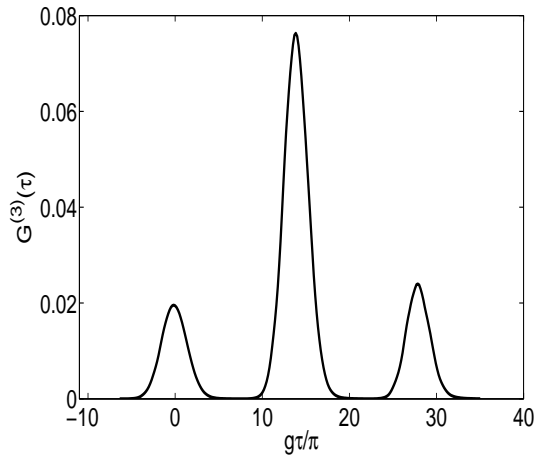}
\caption{The triple coincidence correlation of detecting one
photon at output of each interferometer and one from the input pulse $\Omega_1$ for $T=14\pi/g$. The other parameters are the same as in Fig.~2} \label{fig3}
\end{figure}

In Fig.~\ref{fig3}, we plot
$G^{(3)}(\tau)$ for the same  parameters used in Fig.~\ref{fig2}, where the
time-dependent populations were shown. The computed value
of $G^{(3)}(\tau)$ has three peaks centered at $\tau=0$, $T$, and
$2T$. The first peak at $\tau=0$ correspond to the photons
generated in the early time-bin that have passed through the short arms of the
interferometers. Similarly the peak centered at $\tau=2T$
corresponds to the photon pair generated in the late time-bin that have passed
through the long arms of the interferometers. The central peak at
$\tau=T$ corresponds to the overlap of the photons generated in early
time bin and passed through the longer arms in the interferometers
and the photons generated in the late time-bin and passed through
the short arms. Thus only the central peak contains the information
about the entanglement and can be easily selected by choosing a narrow
time window around $\tau=T$. We have also found that the required value of $T$ is slightly less than the actual time between the pump pulses, which shows that in STIRAP, the  photons are actually
generated before the pump pulse reaches its maximum. For the parameters used in Fig.~\ref{fig2}, the time between pump pulses is $15\pi/g$, but the central peak in Fig.~\ref{fig3} is a maximum for $T=14\pi/g$.
\begin{figure}
\vspace{-0.4cm} \centering
\includegraphics[width=2.5in,height=5.0cm]{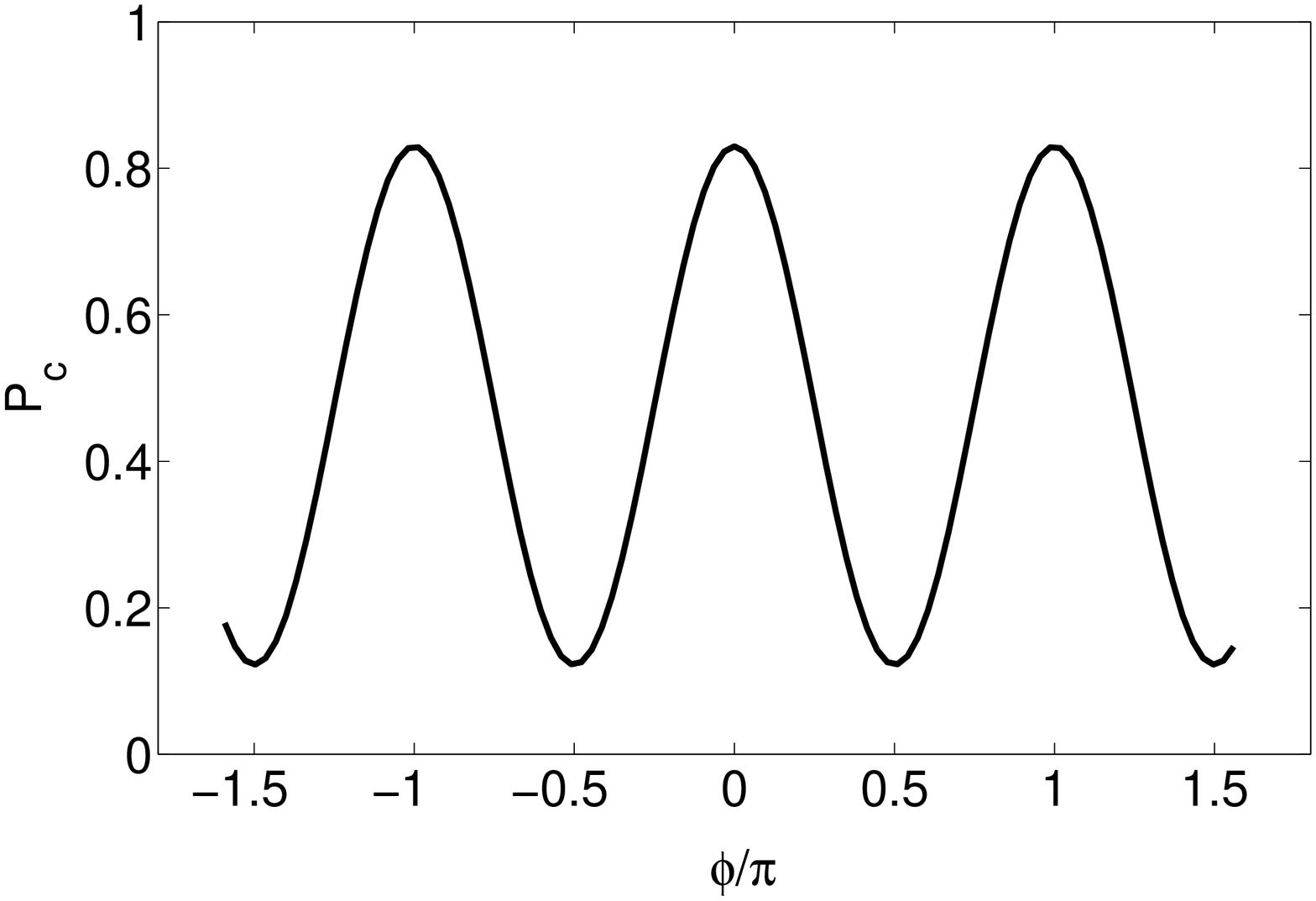}
\caption{The integrated value of the triple coincidence correlation $G^{(3)}(\tau)$ along the central peak at $\tau=T$. The interference pattern appears on changing the phase $\phi$ produced by the interferometers.} \label{fig4}
\end{figure}
\begin{figure}
\centering
\includegraphics[width=2.5in,height=5.0cm]{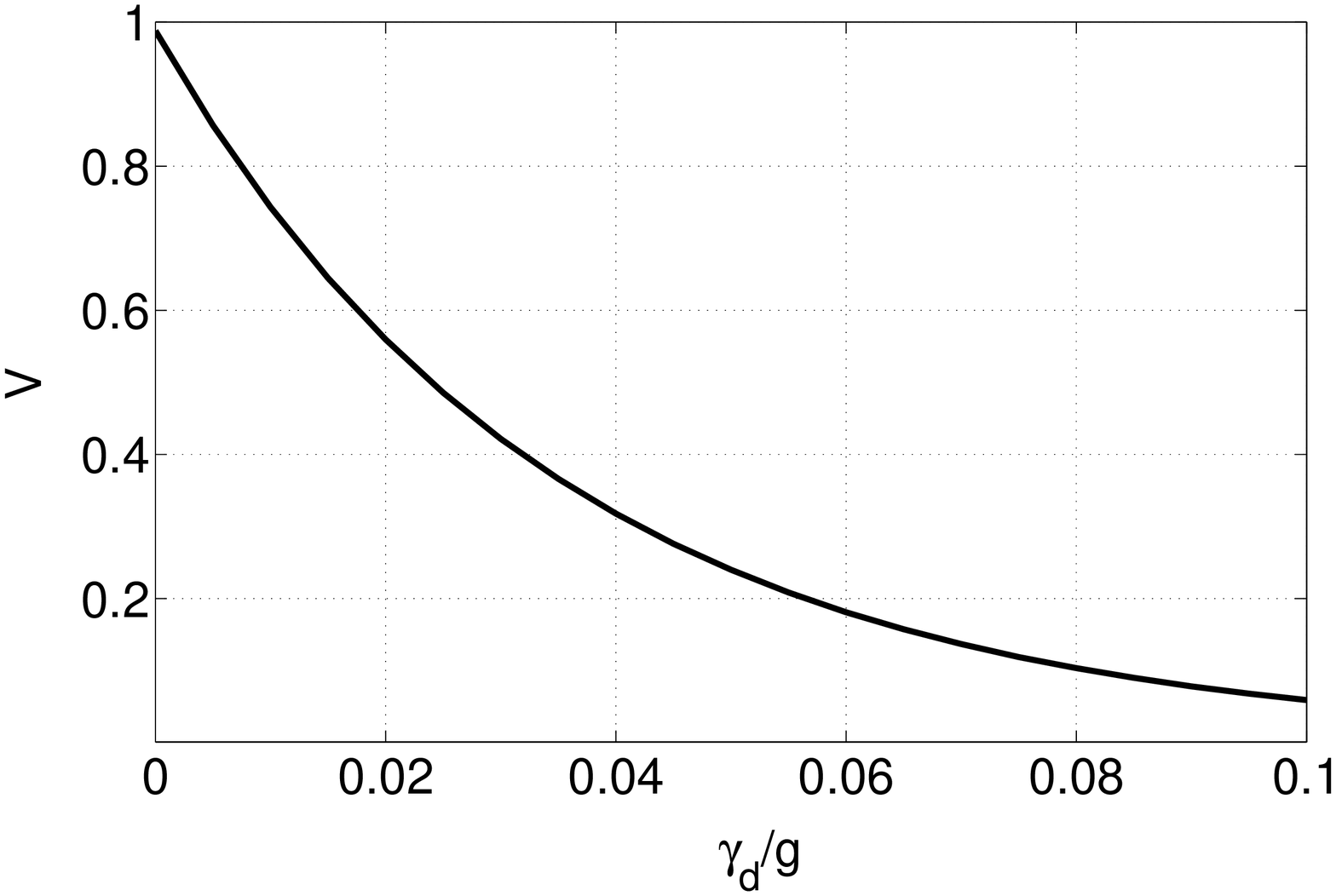}
\caption{The dependence of the visibility, i.e. entanglement of the generated time-bin entangled state of the photon
on dephasing rate $\gamma_d$.} \label{fig5}
\end{figure}

The coherence in the generated state (\ref{tstate}) can be measured by varying the
phase $\phi$ between the overlapping amplitudes corresponding to
the early and the late time bins along the central peak. In
Fig.\ref{fig4}, we plot the interference pattern produced in the measurement of $P_c=\int_{T-T_{bin}}^{T+T_{bin}}G^{(3)}(\tau)d\tau$. The visibility
of the interference pattern, defined as $V$=(maximum of $P_c$ - minimum of $P_c$)/ (maximum of $P_c$ + minimum of $P_c$), gives the concurrence and the purity of the
generated state (\ref{tstate}).

In Fig.~\ref{fig5}, we show the dependence of the visibility on the dephasing
rate. In the presence of pure  dephasing, the visibility,
i.e. concurrence of the generated state (Eq.~\ref{tstate}), is strongly
inhibited.  Further the dephasing in the state (Eq.~\ref{tstate}) only occurs during the
interaction of the pump pulses. The dephasing during the time gap
between the pump pulses plays no role as the coherence in the state is
produced by the coherence between the input pulses. The inhibition
of interference due to dephasing can be understood as due to
pure dephasing (e.g., phonon) interactions the information about the photon generating in different
time bin is partially imprinted in the phonon baths and thus as a result
of quantum complementarity the interference is inhibited. However, for small dephasing rate
$\gamma_d\approx0.01g_{1}$, the value of visibility is larger than
$1/\sqrt{2}$, which is required for violation of Bell's
inequalities\cite{bell}.

\section{Conclusions}
\label{Sec:Conclusions}
We have presented a cavity-QED STIRAP scheme for generating a scalable source of time-bin entangled photon pairs,  and we also investigated the role of pure dephasing on entanglement.
The generated state of the photons can be detected by measuring the correlations between the pump and the generated photons. We found that for small values of pure dephasings, it is possible to achieve larger values of entanglement using current working technologies.

\section{Acknowledgements}
This work was supported by the National Sciences and Engineering
Research Council of Canada, and the Canadian Foundation for
Innovation. We acknowledge useful discussions with Robin Williams and
Gregor Weihs.
\appendix*
\section{Calculation of multi-time correlations}
Here we briefly discuss the method for calculating two-time correlation and four-time correlation used in Sec.~III. We follow the approach discussed by Gardiner and Zoller \cite{book} for evaluating multi-time correlations. The required two time correlation can be expressed as
\begin{eqnarray}
&&\langle
a^{\dag}_1(t)a^{\dag}_2(t+\tau)a_2(t+\tau)a_1(t)\rangle\nonumber\\
&&=Tr\left\{a_2(t+\tau)a_1(t)\rho(0) a^{\dag}_1(t)a^{\dag}_2(t+\tau)\right\}\\
&&=Tr\left\{a_2(t+\tau)\rho'(t)a^{\dag}_2(t+\tau)\right\},\\
&&=Tr\left\{a_2\rho'(t+\tau)a^{\dag}_2\right\}
\label{2time}
\end{eqnarray}
where $Tr$ stands for trace and operators $a_i$ appearing without time parenthesis are in Schrodinger picture, $\rho'(t)=a_1(t)\rho(0) a^{\dag}_1(t)=a_1\rho(t) a^{\dag}_1$ is calculated after evolving
the initial state $\rho(0)=\rho_{mm}|m\rangle\langle m|$ for time $t$ using Eqs.(\ref{meq1})-(\ref{meq2}) and then operating by $a_1$ and $a^{\dag}_1$ from left and right, respectively. Clearly, $\rho'$ also follows the same equations of motions (\ref{meq1})-(\ref{meq2}). Now, using initial value $\rho'(t)=a_1\rho(t) a^{\dag}_1$ at time $t$, and evolving for time $\tau$, $\rho'(t+\tau)$ is calculated. The value of the required correlation is calculated using Eq. (\ref{2time}). A similar straightforward approach, considering the times appearing in the $a$ operators in ascending order, is applied in evaluating the four time correlations as follows:
\begin{eqnarray}
&&\langle
a^{\dag}_1(t)a^{\dag}_2(t+\tau)a_2(t-T+\tau)a_1(t-T)\rangle, \nonumber\\
&&=Tr\left\{a_2(t-T+\tau)a_1(t-T)\rho(0) a^{\dag}_1(t)a^{\dag}_2(t+\tau)\right\} \nonumber \\
&& \\
&&=Tr\left\{a_2(t-T+\tau)\rho_1(t-T)a^{\dag}_1(t)a^{\dag}_2(t+\tau)\right\},\\
&&=Tr\left\{\rho_2(t-T+\tau)a^{\dag}_1(t)a^{\dag}_2(t+\tau)\right\}, \\
&&=Tr\left\{\rho_3(t)a^{\dag}_2(t+\tau)\right\}, \\
&&=Tr\left\{\rho_3(t+\tau)a^{\dag}_2\right\},
\label{4times}
\end{eqnarray}
where $\rho(0)=\rho_{mm}|m\rangle\langle m|$, $\rho_1(t-T)\equiv a_1(t-T)\rho(0)\equiv a_1\rho(t-T)$, $\rho_2(t-T+\tau)\equiv a_2(t-T+\tau)\rho_1(t-T)\equiv a_2\rho_1(t-T+\tau)$, and $\rho_3(t)\equiv\rho_2(t-T+\tau)a_1^{\dag}(t)\equiv\rho_2(t)a_1^{\dag}$. Thus the density matrices $\rho$, $\rho_1$, $\rho_2$, and $\rho_3$ are evolved for times, $0$ to $t-T$, $t-T$ to $t-T+\tau$, $t-T+\tau$ to $t$, and $t$ to $t+\tau$ respectively. The evolution of $\rho(0)$ is given by Eqs. (\ref{meq1})-(\ref{meq2}), while the evolutions of density matrices $\rho_i$ for i=1,2,3, follow the similar equations, written for $\rho$, as 
\begin{eqnarray}
\label{meqcorr1}
\dot\rho_{Yy}&=&-ig_1\rho_{uy}-ig_2\sqrt{2}\rho_{G'y}+ig_2\rho_{YG}\nonumber\\
&&-(\gamma_2+\kappa/2+\gamma_d)\rho_{Yy},\\
\dot\rho_{G'G}&=&-ig_2\sqrt{2}\rho_{YG}+ig_2\rho_{G'y}-\frac{3}{2}\kappa\rho_{G'G},\\
\dot\rho_{Gg}&=&-ig_2\rho_{yg}-\frac{1}{2}\kappa\rho_{Gg},\\
\dot\rho_{uy}&=&-i\Delta_1\rho_{uy}-i\Omega_p(t)\rho_{my}-ig_1\rho_{Yy}+ig_2\rho_{uG}\nonumber\\
&&-(\gamma_1+\gamma_2/2+3\gamma_d/2)\rho_{uy},\\
\dot\rho_{G'y}&=&i\Delta_2\rho_{G'y}-ig_2\sqrt{2}\rho_{Yy}+ig_2\rho_{G'G}\nonumber\\
&&-(\kappa+\gamma_2/2+\gamma_d/2)\rho_{G'y},\\
\dot\rho_{YG}&=&-i\Delta_2\rho_{YG}-ig_1\rho_{uG}-ig_2\sqrt{2}\rho_{G'G}+ig_2\rho_{Yy}\nonumber\\
&&-(\kappa+\gamma_2/2+\gamma_d/2)\rho_{YG},\\
\dot\rho_{yg}&=&-i\Delta_2\rho_{yg}-ig_2\rho_{Gg}-\frac{1}{2}(\gamma_2+\gamma_d)\rho_{yg},\\
\dot\rho_{my}&=&-i(\Delta_1-\Delta_p)\rho_{my}-i\Omega_p^*(t)\rho_{uy}+ig_2\rho_{mG}\nonumber\\
&&-(\gamma_2/2+\gamma_d)\rho_{my},\\
\dot\rho_{uG}&=&-i(\Delta_1+\Delta_2)\rho_{uG}-i\Omega_p\rho_{mG}-ig_1\rho_{YG}+ig_2\rho_{uy}\nonumber\\
&&-(\kappa/2+\gamma_1+\gamma_d)\rho_{uG},\\
\dot\rho_{mG}&=&-i(\Delta_1+\Delta_2-\Delta_p)\rho_{mG}-i\Omega_p^*(t)\rho_{uG}+ig_2\rho_{my}\nonumber\\
&&-\frac{1}{2}(\kappa+\gamma_d)\rho_{mG},\\
\dot\rho_{G'g}&=&-ig_2\sqrt{2}\rho_{Yg}-\kappa\rho_{G'g},\\
\dot\rho_{Yg}&=&-i\Delta_2\rho_{Yg}-ig_2\sqrt{2}\rho_{G'g}-ig_1\rho_{ug}\nonumber\\
&&-\frac{1}{2}(\kappa+\gamma_2+\gamma_d)\rho_{Yg},\\
\dot\rho_{ug}&=&-i(\Delta_1+\Delta_2)\rho_{ug}-i\Omega_p\rho_{mg}-ig_1\rho_{Yg}\nonumber\\
&&-(\gamma_1+\gamma_d)\rho_{ug},\\
\dot\rho_{mg}&=&-i(\Delta_1+\Delta_2-\Delta_p)\rho_{mg}-i\Omega_p^*(t)\rho_{ug} \nonumber \\
&&  -\frac{1}{2}\gamma_d\rho_{mg}.
\label{meqcorr2}
\end{eqnarray}
Finally, the value of four-times correlations used in Sec.III are found using Eq.~(\ref{4times}).

\end{document}